\documentstyle[11pt,aaspp]{article}
\def\t0{\theta_{\circ}}

\def\be{\begin{equation}}
\def\en{\end{equation}}

\def\msun{M_{\sun}}

\begin{document}
\title {Mid-Infrared Imaging of Candidate Vega-Like Systems}
\author{Ray Jayawardhana\altaffilmark{1,2,3},
R. Scott Fisher\altaffilmark{2,3,4},
Charles M. Telesco\altaffilmark{2,3,4},
Robert K. Pi\~na\altaffilmark{2,3,4},\\
David Barrado y Navascu\'es\altaffilmark{5},
Lee W. Hartmann\altaffilmark{6},
and Giovanni G. Fazio\altaffilmark{6}}
\altaffiltext{1}{Department of Astronomy, University of California, Berkeley, 
601 Campbell Hall, Berkeley, CA 94720; Electronic mail: 
rayjay@astro.berkeley.edu}
\altaffiltext{2} {Visiting Astronomer, Cerro Tololo Interamerican Observatory,
National Optical Astronomy Observatories, which is operated by the Association
of Universities for Research in Astronomy, Inc. (AURA) under cooperative
agreement with the National Science Foundation.}
\altaffiltext{3} {Visiting Astronomer, W.M. Keck Observatory, which is
operated as a scientific partnership among the California Institute of
Technology, the University of California, and the National Aeronautics
and Space Administration.}
\altaffiltext{4}{Department of Astronomy, University of Florida, Gainesville, FL 32611}
\altaffiltext{5}{Departamento de F\'{\i}sica Te\'orica, C-XI.
Universidad Aut\'onoma de Madrid, E-28049 Madrid, Spain}
\altaffiltext{6}{Harvard-Smithsonian Center for Astrophysics, 60 Garden St., Cambridge, MA 02138}

\begin{abstract}
We have conducted deep mid-infrared imaging of a relatively nearby sample of 
candidate Vega-like stars using the OSCIR instrument on the CTIO 4-meter 
and Keck II 10-meter telescopes. Our discovery of a spatially-resolved disk 
around HR 4796A has already been reported (Jayawardhana et al. 1998). Here we 
present imaging observations of the other members of the sample, including the 
discovery that only the primary in the HD 35187 binary system appears to 
harbor a substantial circumstellar disk and the possible detection of 
extended disk emission around 49 Ceti. 

We derive global properties of the 
dust disks, place constraints on their sizes, and discuss several interesting 
cases in detail. Although our targets are believed to be main sequence stars, 
we note that several have large infrared excesses compared to prototype 
Vega-like systems, and may therefore be somewhat younger. The disk size
constraints we derive, in many cases, imply emission from relatively large 
($\gtrsim$ 10$\mu$m) particles at mid-infrared wavelengths.
\end{abstract}

\keywords{planetary systems --circumstellar matter --stars: 
binaries -- stars: pre-main-sequence --stars: early-type}

\section{Introduction}
Early in its mission, the {\it InfraRed Astronomy Satellite (IRAS)} detected
thermal emission from solid grains with temperatures of 50-125 K and 
fractional luminosities ($L_{dust}/L_{*}$) in the range 
$10^{-5}$-$10^{-3}$ around four main sequence stars: Vega, Fomalhaut, 
$\beta$ Pictoris, and $\epsilon$
Eridani. Coronagraphic observations of $\beta$ Pic revealed that the grains
do indeed lie in a disk, perhaps associated with a young planetary
system (Smith \& Terrile 1984). Subsequent surveys of {\it IRAS} data
revealed over 100 other main sequence stars of all spectral classes with 
far-infrared excesses indicative of circumstellar disks (Aumann 1985; 
Sadakane \& Nishida 1986; Walker \& Wolstencroft 1988;
Jascheck et al. 1991; Oudmaijer et al. 1992; Cheng et al. 1992; Backman 
\& Paresce 1993; Mannings \& Barlow 1998). 

In most ``Vega-like'' systems, the dust grains responsible for the infrared 
emission are thought to be continually replenished by 
collisions and sublimation of larger bodies, because the timescales for grain
destruction by Poynting-Robertson (PR) drag and ice sublimation are much
shorter than the stellar main sequence lifetimes (Nakano 1988; 
Backman \& Paresce 1993). In other words, the disks around main sequence 
stars are likely to be debris disks rather than protoplanetary 
structures. These optically thin debris disks contain much less dust (and gas)
than the massive optically thick structures observed around young 
pre-main-sequence stars (e.g., Strom et al. 1989; Beckwith et al. 1990). 
Recent high-resolution sub-millimeter images of the four prototype debris 
disks confirm the presence of inner disk cavities, which may persist due to 
planets within the central void consuming or perturbing grains inbound under 
the influence of the PR drag (Holland et al. 1998; Greaves et al. 1998). 

Over the past two years, we have conducted deep mid-infrared imaging of
a relatively nearby sample of candidate Vega-like stars. Our goals are to 
study the diversity of debris disks and to explore possible evolutionary
effects. Imaging at 18$\mu$m and 10$\mu$m with an angular resolution of 1'' 
or better is extremely valuable for confirming {\it IRAS}-detected excesses,
constraining disk global properties, and estimating the typical grain sizes 
from measurements of the emission temperature and scale (e.g., Jura et al. 
1993, 1995).

Our discovery of a spatially-resolved disk around HR 4796A has already been 
reported (Jayawardhana et al. 1998; Telesco et al. 2000). Its low-mass binary 
companion, at a projected distance of 500AU, is estimated to be $8\pm3$ Myrs 
old, an age comparable to the timescale expected for planet formation (Strom, 
Edwards, \& Skrutskie 1993; Podosek \& Cassen 1994). Interestingly, the 
HR 4796A disk does have an inner cavity of solar system dimensions, and the 
disk dust mass is estimated to be only $\sim$1M$_{Earth}$. It is likely that 
circumstellar disks evolve from masssive, optically thick, actively-accreting 
structures to low-mass optically thin structures with inner holes in about 10 
Myrs, and that the disk evolution is closely linked to planet formation
(Jayawardhana et al. 1999a, 1999b, 2001; Jayawardhana 2000 and references 
therein).

Here we present imaging observations of the other members of the sample, 
including the discovery that only the primary in the HD 35187 binary system 
appears to harbor a substantial circumstellar disk and the possible detection 
of extended disk emission around 49 Ceti. We derive global properties of 
the dust disks, place constraints on their sizes, and discuss several 
interesting cases in detail.

\section{Observations}
Our sample consists of 11 nearby main sequence stars with known
{\it IRAS} excess at mid- and/or far-infrared wavelengths. The primary
criteria for selection were proximity, spectral type (biased towards 
early types), and brightness in the mid-infrared. Table 1 lists the
adopted stellar properties for the 11 objects. 

During three observing runs in 1999, we have obtained deep mid-infrared 
images of the candidate stars using the OSCIR instrument on the 4-meter 
Blanco telescope at Cerro Tololo Interamerican Observatory (CTIO) and the 
10-meter Keck II telescope. The log of our observations is given in Table 
2. OSCIR is a mid-infrared imager/spectrometer built at the University of 
Florida\footnote[1]{Additional information on  OSCIR is available on the 
Internet at www.astro.ufl.edu/iag/.}, using a 128$\times$128 Si:As Blocked 
Impurity Band (BIB) detector developed by Boeing. 

On the CTIO 4-m telescope, OSCIR has a plate scale of 0.183''/pixel, which 
gives a field of view of 23''$\times$23''. Our observations were made using 
the standard chop/nod technique with a chop frequency of 5 Hz and a throw of 
23'' in declination. On Keck II, its plate scale is 0.062''/pixel, providing 
a 7.9''$\times$7.9'' field of view. Here we used a chop frequency of 4 Hz and 
a throw of 8''. Images were obtained in the N(10.8 $\mu$m) band for 10 of
the 11 targets in our sample, and in the IHW18(18.2 $\mu$m) band for 9 objects.

\section{Results}
In Table 3, we present the measured fluxes in the N and IHW18 bands for
the entire sample. Within the 10\% errors, our measurements are consistent
with published ground-based near-infrared and {\it IRAS} 12$\mu$m and 25$\mu$m 
fluxes, ruling out the possibility that {\it IRAS}-detected excesses are due 
to contamination from unrelated sources within the large {\it IRAS} beam. None 
of the objects clearly shows spatially-resolved emission in our images. 

Figure 1 presents normalized scans through our targets and the PSF stars;
in cases where the target and/or the PSF appeared asymmetric, the scan
shown is along the ``major axis''. Most targets are indistinguishable
from point sources, except possibly 49 Ceti, HD 169142, and HD 135344 
(at 18$\mu$m). None of the three is obviously elonagted, and particularly
in the case of HD 169142 and HD 135344, the larger full-width at half-maximum
may be the result of seeing effects, because the 
integration time on the source was {\it much} longer than that 
on the PSF star. We discuss 49 Ceti in some detail in Section 4.2.

We can use comparison with the PSF star to constrain the radius of the 
mid-infrared emitting region in each target. In addition, we also conducted 
the following experiment to derive a limit on the mid-infrared disk size: 
an image of a simple disk model with given inner and outer radii was convolved
with the PSF, and added to an appropriately scaled point-source. The flux
in the point source and the disk were determined from the observed stellar
and excess emission in the N and IHW18 bands. By varying the
outer radius, we were able to constrain the maximum radius the disk could
have without appearing to be more extended than the PSF star. Using this
test as well as direct inspection of scans shown in Figure 1, we obtained
constraints on the disk radii listed in column 6 of Table 4. 

\section{Discussion}
We derived mid-infrared excesses of our sample by subtracting the expected 
photospheric flux assuming V-N and V-[12] colors given by Kenyon \& Hartmann 
(1995). Then, following Backman \& Gillett (1987), we can write the 
fractional luminosity of dust as $\tau = L_{dust}/L_{*}$, where $L_{dust}$ 
and $L_{*}$ are the luminosities of the dust and the star, respectively. The 
$\tau$ values derived for our sample, based on long-wavelength excesses, range 
from $\sim 10^{-6}$ to $\sim 10^{-1}$ (see Table 4). For a flat optically 
thick disk, the maximum value of $\tau$ is 0.25 while for a `flared' disk it 
could be roughly 0.5 (Kenyon \& Hartmann 1987). The prototype Vega-like stars 
have $\tau = 10^{-5} - 10^{-3}$, implying low optical depths at all 
wavelengths. A couple of stars in our sample have $\tau$ larger than the flat 
disk maximum. Such large infrared excesses require either an additional source
of infrared emission or an alternate geometry for the dust (Sylvester et 
al. 1996). In fact, the fractional excess luminosities of several stars
in our sample are similar to those of Herbig Ae/Be stars and T Tauri stars
(Cohen et al. 1989; Hillenbrand et al. 1992), suggesting that they may be at 
an evolutionary stage not far removed from these pre-main-sequence objects.
(See Section 4.1 for a more detailed discussion of one example, HD 35187.)
%Indeed, Song et al. (2000) find that there is a bimodal distribution of 
%$\tau$, centered at 0.0003 and 0.1, and suggest a possible evolutionary 
%sequence for Vega-like systems.

The long-wavelength excesses of all our targets can be fit 
fairly well by a single-temperature blackbody; the dust temperatures derived 
from such fits are listed in column 3 of Table 4. A single-temperature fit 
implying a single radial location for the material is of course only schematic.
Many combinations of radial density and grain size distributions can produce 
an SED which resembles portions of a Planck curve (Backman \& Paresce 1993).
Nevertheless, it is interesting to note that all but one of the objects have 
a dominant dust component with derived temperatures in the 70-150 K range. The 
exception is 51 Oph which appears to harbor primarily hot circumstellar dust 
with a temperature of $\gtrsim$500 K; we discuss it in more detail in Section 
4.3. Several of the other systems also require an additional warmer component 
to account for all of the excess at $\lambda < 12 \mu$m. The SEDs for two such 
cases are shown in Figure 2.

We can use our data in combination with the {\it IRAS} measurements to 
constrain the location and size of the dust particles in these Vega-like
systems. If the grain emissivity varies as $\nu^{p}$, where $p$ = 0 
corresponds to blackbody emission from large grains and $p$ = 1 corresponds 
to emission from small grains, then following Jura et al. (1993), we can 
write that at a distance from the star $D$, the grains reach a temperature 
$T_{gr}$, given by the relationship

\begin{equation}
D = 0.5 R_{*} \left (T_{*}/T_{gr} \right )^{(2+0.5p)},
\end{equation}
where $R_{*}$ and $T_{*}$ are the radius and temperature of the central star,
respectively. This relation ignores dust scattering. We list the derived disk 
radii assuming $p$ = 0 and $p$ = 1 in column 4 and 5 of Table 4. The radius
limits from our imaging, listed in column 6, rule out $p$ = 1 grains in most 
cases, and are consistent with the hypothesis that the grains act as 
blackbodies at wavelengths $\lesssim$ 60 $\mu$m. However, realistic models
of debris disks must include a distribution of grain sizes and compositions.
Detailed modeling of the HR 4796A disk, for example, implies 2-3$\mu$m grains
(Telesco et al. 2000) as well as a population of much larger particles
(Augereau et al. 1999). In $\beta$ Pic, the spectral form of the silicate
feature suggests the presence of 2-3$\mu$m grains (e.g., Aitken et al. 1993)
while far-infrared/sub-millimeter fluxes provide evidence for larger grains
(e.g., Zuckerman \& Becklin 1993).

Notes on three particularly interesting objects follow.

\subsection{HD 35187}
HD 35187 (SAO 77144) is an A2/A7 binary at a distance of 150$\pm$55 pc
in Taurus. The {\it Hipparcos} catalog has
(surprisingly) assigned the identifier `B' to the A2 star, which is
the brighter, more northerly component; Dunkin \& Crawford (1998)
followed the same nomenclature and we will too, in order to avoid
confusion. The two stars are separated by 1.39'' with a position angle of
192 degrees, and have similar V magnitudes ($V_B=8.6$, $V_A=8.7$). 

Our near- and mid-infrared images make it possible for the first time to 
directly determine which star in the pair harbors the excess emission 
detected by {\it IRAS}. In the near-infrared, where the radiation is primarily 
photospheric, HD 35187B and HD 35187A have a flux ratio of $\sim$1.6. At 
4.8$\mu$m, 10.8$\mu$m, and 18.2$\mu$m, the primary becomes increasingly 
dominant over the secondary, suggesting that most of the circumstellar dust 
in the system resides around HD 35187B. 

This result is consistent with that of Dunkin \& Crawford (1998) who 
obtained spatially resolved optical spectra of the individual stars. They 
detected net H$\alpha$ emission, circumstellar absorption lines and 
significant circumstellar extinction for HD 35187B only, and first
suggested that the observed infrared excess is due to a disk surrounding
the primary (B) alone.

Such ``mixed'' binary systems, where one star appears to have a circumstellar
disk but the other does not, are rare among low-mass pre-main sequence
binaries (Prato \& Simon 1997). However, a few examples have been found,
particularly in systems which may be in transition from actively accreting
classical T Tauri stage to passive remnant disks (e.g., Jayawardhana et al.
1999a). 

The present disk configuration in HD 35187 could be either due to the
details of binary formation or to the evolutionary histories of disks.
Smoothed particle hydrodynamics (SPH) simulations by Bate \& Bonnell
(1997) show that the fractions of infalling material that are captured by
each protostar and the fraction which forms a circumbinary disk depend on
the binary's mass ratio and the parent cloud's specific angular momentum,
$j_{infall}$. For accretion with low $j_{infall}$, most of the infalling
material is captured by the primary. For gas with intermediate
$j_{infall}$, the fraction captured by the primary decreases and that
captured by the secondary increases. For 
higher $j_{infall}$, more and more gas goes into a circumbinary disk instead 
of circumstellar disks. Thus, it could be that infall from a low $j_{infall}$ 
cloud led to a more massive disk around the primary in HD 35187. However, 
given their roughly equal masses (2.1$\msun$ vs. 1.7$\msun$; Dunkin \& 
Crawford 1998), it is not clear why one star would capture much more material 
than the other. One possibility is that as protostars, the two components had 
very different masses, with what is now the primary accumulating most of its 
mass at the end of accretion. Another possibility is that there is a very 
low-mass, so far undetected, close companion around HD 35187A which 
accelerated the depletion of its disk.

There is some ambiguity about the evolutionary status of HD 35187. It has 
been classified as a Herbig Ae pair by some authors (B\"ohm \& Catala 1995; 
Grady et al. 1996) and as a Vega-like system by others (Walker \& Wolstencroft 
1988; Sylvester et al. 1996). Both components of HD 35187 satisfy most of
the criteria for Herbig Ae/Be stars, which are considered to be the 
higher-mass counterparts to young T Tauri stars. The observed fractional 
infrared luminosity ($L_{IR}/L_{*}$) of 0.16 is also comparable to those
of other well known Herbig Ae/Be stars. On the other hand, both stars in the 
HD 35187 system lie very close to the zero-age main sequence on the H-R 
diagram, at a {\it Hipparcos} distance of 150 pc (Dunkin \& Crawford 1998).
Furthermore, the SED of the system (Sylvester et al. 1996) reveals a dip near 
10$\mu$m similar to that reported by Waelkens, Bogaert \& Walters (1994) for 
their sample of evolved Herbig Ae/Be stars. (Our flux measurements confirm
a dip at 10$\mu$m in the HD 35187 system: our N-band flux is only 3.4 $\pm$ 
10\% Jy compared to the {\it IRAS} 12$\mu$m flux of 5.39 Jy.) Thus, HD 35187 
is likely to be at an intermediate age between young Herbig Ae/Be stars and 
older `classical' Vega-like sources. This conclusion is consistent with its
apparent age of 10 Myr, based on H-R diagram evolutionary tracks (Dunkin
\& Crawford 1998), and places it at an interesting epoch in disk
evolution (see Jayawardhana et al. 1999b).

\subsection{49 Ceti}
49 Ceti was first identified as a candidate Vega-like source by Sadakane
\& Nishida (1986) when they searched the {\it IRAS} point-source catalog
for {\it Bright Star Catalog (BSC)} members with significant 60$\mu$m excess.
As Jura et al. (1993) have pointed out, of the $\sim$1500 A-type stars in 
the {\it BSC} only three have $L_{IR}/L_{*} > 10^{-3}$. 49 Ceti is one, 
and the other two are $\beta$ Pictoris and HR 4796A, both of which have 
spatially resolved debris disks. Given its illustrious company and its
relative proximity at 61 pc, 49 Ceti is naturally a prime target for 
any disk search. However, the low flux levels in the mid-infrared 
($F12=0.33$ Jy, $F25 < 0.41$ Jy) make it a challenge to detect an 
extended disk; for example, if the observed N-band excess were distributed 
in a 100AU-radius disk, its surface brightness would be less than 2.5 mJy
per square arcsecond, allowing at best a $\sim$1-$\sigma$ detection in 
our 120 sec observation. 

Our N-band image of 49 Ceti does not exhibit any obvious elongation, but 
there is tentative evidence for extended emission. Figure 1 shows 
normalized scans through 49 Ceti and the PSF star. The difference in the
width of the source and the PSF star suggest that emission from 49 Ceti
may be marginally resolved. It is possible that the effect is due to
seeing. However, none of the other targets observed on the
same night shows a similar effect (Fisher 2001). The observed source
size is consistent with a $r\sim$50 AU (i.e., $p \sim 0$) disk (See Table
4). 

Recently, Guild et al. (1999) have independently detected low-level extended 
emission at 18$\mu$m around 49 Ceti. They derive an angular size of 1.5'',
consistent with our 10$\mu$m observation, and a position angle of 40 degrees,
which is not well constrained by our data.

\subsection{51 Oph}
51 Oph is a B9.5Ve star with an unusually large mid-infrared excess; in fact, 
its IRAS 12$\mu$m excess is 3.9mag compared with 0.96mag for $\beta$ Pic 
(Cot\'e \& Waters 1987; Waters et al. 1988). Based on similarities between 
the optical and ultraviolet lines from circumstellar gas around $\beta$ Pic 
and 51 Oph, Grady \& Silvis (1993) have suggested that the disk around 51 Oph 
should also be seen edge-on. However, 51 Oph is now known to be 6.8 times 
further away than $\beta$ Pic (i.e., 131 vs 19.3 pc), and there is no 
definitive evidence yet that 51 Oph's circumstellar dust is in a disk rather 
than a shell. 

No extended disk is detected in our deep 18$\mu$m image of 51 Oph at Keck,
consistent with the point-like appearance at 10$\mu$m in observations
reported by Lagage \& Pantin (1994). Our data indicate that the surface
brightness 1'' from the center in all directions would have to be less than 
2\% of the peak flux. Mid-infrared spectroscopy has revealed that 
the 10$\mu$m excess is in 51 Oph is resolved into a prominent, broad silicate 
emission feature that would seem to be associated with hot ($\gtrsim$500 K), 
small dust particles (Fajardo-Acosta et al. 1993; Lynch et al. 1994). Our
imaging observations, combined with the fact that the {\it IRAS} SED drops
rapidly beyond 25$\mu$m, strengthen the case for hot dust in close
proximity ($r \lesssim$ 5 AU) to the star rather than a $\beta$ Pic-like
disk extending to well over 100 AU. Recently van den Ancker et al. (2001) have 
detected hot molecular gas and partially crystalline silicate dust in a 
spectrum of 51 Oph obtained by the Infrared Space Observatory (ISO), again
highly unusual for a Vega-like system. One possibility, as van den Ancker et 
al. point out, is that 51 Oph is a Be star which underwent a recent episode
of mass loss.

\section{Summary}
We have carried out deep mid-infrared imaging observations of a sample
of nearby main sequence stars with {\it IRAS}-detected excess emission.
We clearly resolve an extended disk in only one case (Jayawardhana et al.
1998). For 11 other systems, we derive constraints on the global properties
of presumed dust disks, including fractional dust luminosity, characteristic 
temperature, disk size, and grain emissivity. Several systems in our sample
have large infrared excesses compared to prototype Vega-like systems, and 
may therefore be somewhat younger. 51 Oph is unusual in having primarily
a hot ($\gtrsim$500 K) dust component. In the A2/A7 binary HD 35187, we find 
that only the primary harbors substantial dust emission. We may have 
marginally resolved the disk around 49 Ceti, but more sensitive observations 
are required for confirmation.

\bigskip
We wish to thank the staff of CTIO and Keck Observatory for their 
outstanding support. RJ holds a Miller Research Fellowship at the 
University of California, Berkeley. This work was supported by the 
Smithsonian Institution, NASA (through a grant to RJ administered by 
the AAS) and NSF (through a grant to the University of Florida). Our 
research has made use of ADS and SIMBAD databases.

\newpage
 
\centerline{\bf Figure Captions}

Figure 1. Scans through the target sources and the PSF stars. In all plots,
        the target is represented by the solid line and the PSF star by the 
        dashed line. 

Figure 2. Spectral energy distributions of HD 143006 ({\it left}) and
        HD 135344 ({\it right}) with single temperature blackbody fits
        to the photospheric emission and the far-infrared excess. 

\clearpage
\begin{table}
\begin{scriptsize}
\begin{center}
\renewcommand{\arraystretch}{1.2}
\begin{tabular}{lccccccc}
\multicolumn{8}{c}{\scriptsize TABLE 1}\\
\multicolumn{8}{c}{\scriptsize OBSERVATIONAL PROPERTIES AND ADOPTED
PARAMETERS OF TARGET STARS\tablenotemark{a}}\\
\hline
\hline
HD & HR & Other name & Sp. Type & V & Distance (pc) & $T _*$ (K) & $R_*$ (cm)\\
\hline
9672 & 451 & 49 Ceti & A1V & 5.62 & 61 & 9230 & $1.6\times10^{11}$\\
35187A\tablenotemark{b} & & & A7V & 8.73 & 150 & 7850 & $1.1\times10^{11}$\\
35187B\tablenotemark{b} & & & A2V & 8.59 & 150 & 8970 & $1.5\times10^{11}$\\
38678 & 1998 & 14 Lep & A2V & 3.55 & 21 & 8970 & $1.5\times10^{11}$\\
74956 & 3485 & $\delta$ Vel & A1V & 1.95 & 24 & 9230 & $1.6\times10^{11}$\\
102647 & 4534 & $\beta$ Leo & A3V & 2.14 & 11 & 8720 & $1.4\times10^{11}$\\
135344\tablenotemark{c} & & & F8V & 8.63 & 84 & 6200 & $8.0\times10^{10}$\\
143006\tablenotemark{c} & & BD-22 4059& G5V & 10.18 & 82 & 5770 & $6.4\times10^{10}$\\
155826 & 6398 & & F7V & 5.96 & 31 & 6280 & $8.3\times10^{10}$\\
158643 & 6519 & 51 Oph & B9.5V & 4.81 & 131 & 10000 & $1.7\times10^{11}$\\
163296 & & Hen 3-1524& A1V & 6.87 & 122 & 9230 & $1.6\times10^{11}$\\
169142\tablenotemark{c} & & & A5V & 8.15 & 145 & 8200 & $1.2\times10^{11}$\\
\hline
\end{tabular}
\end{center}
\tablenotetext{a}{\scriptsize Spectral type, V magnitude and distance
are from SIMBAD database, unless otherwise noted. $T _*$ and $R_*$ for 
a given spectral type are based on {\it Allen's Astrophysical Quantities},
2000, ed. A.N. Cox.}
\tablenotetext{b}{\scriptsize Spectral type, V magnitude and distance are
from Dunkin \& Crawford 1998.}
\tablenotetext{c}{\scriptsize Spectral type, V magnitude and distance are
from Sylvester et al. 1996.}
\end{scriptsize}
\end{table}

\clearpage

\begin{table}
\begin{scriptsize}
\begin{center}
\renewcommand{\arraystretch}{1.2}
\begin{tabular}{lccccc}
\multicolumn{6}{c}{\scriptsize TABLE 2}\\
\multicolumn{6}{c}{\scriptsize LOG OF OBSERVATIONS}\\
\hline
\hline
UT Date & Telescope & Target & Filter & On-source integration & Flux standards\\
\hline
1999 Feb 24 & CTIO 4m & HD 74956 & N & 600 sec& $\gamma$ Ret, $\lambda$ Vel\\
1999 Feb 24 & CTIO 4m & HD 74956 & IHW18& 600 sec& $\gamma$ Ret,$\lambda$ Vel\\
1999 Feb 25 & CTIO 4m & HD 38678 & N & 600 sec & $\gamma$ Ret, $\gamma$ Cru\\
1999 Feb 26 & CTIO 4m & HD 35187 & K & 300 sec & $\lambda$ Vel\\
1999 Feb 26 & CTIO 4m & HD 35187 & M & 211 sec & $\lambda$ Vel\\
1999 Feb 26 & CTIO 4m & HD 35187 & N & 150 sec & $\alpha$ Tau, $\lambda$ Vel\\
1999 Feb 26 & CTIO 4m & HD 35187 & IHW18 & 600 sec & $\alpha$ Tau, $\lambda$ Vel\\
1999 Feb 27 & CTIO 4m & HD 102647 & IHW18 & 100 sec & $\alpha$ CMa, $\gamma$ Cru\\
1999 Feb 27 & CTIO 4m & HD 102647 & N & 300 sec & $\alpha$ CMa, $\gamma$ Cru\\
1999 May 3 & Keck II & HD 135344 & IHW18 & 900 sec & $\mu$ UMa, $\alpha$ Boo\\
1999 May 3 & Keck II & HD 135344 & N & 120 sec & $\mu$ UMa, $\alpha$ Boo\\
1999 May 3 & Keck II& HD 143006 & IHW18 & 600 sec & $\alpha$ Boo, $\eta$ Sgr\\
1999 May 3 & Keck II & HD 143006 & N & 600 sec & $\alpha$ Boo, $\eta$ Sgr\\
1999 May 3 & Keck II & HD 155826 & IHW18 & 600 sec & $\alpha$ Boo, $\eta$ Sgr\\
1999 May 3 & Keck II & HD 155826 & N & 60 sec & $\alpha$ Boo, $\eta$ Sgr\\
1999 May 3 & Keck II & HD 163296 & IHW18 & 300 sec& $\eta$ Sgr, $\gamma$ Aql\\
1999 May 3 & Keck II & HD 163296 & N & 300 sec & $\eta$ Sgr, $\gamma$ Aql\\
1999 May 3 & Keck II & 51 Oph & IHW18 & 600 sec & $\eta$ Sgr, $\gamma$ Aql\\
1999 May 4 & Keck II & HD 169142 & N & 600 sec& $\alpha$ Boo, $\gamma$ Aql\\
1999 May 4 & Keck II & HD 169142& IHW18& 600 sec& $\alpha$ Boo, $\gamma$ Aql\\
1999 Nov 20 & Keck II & 49 Ceti & N & 120 sec & $\beta$ Peg, $\alpha$ Ari\\
\hline
\end{tabular}
\end{center}
\end{scriptsize}
\end{table}

\clearpage

\begin{table}
\begin{scriptsize}
\begin{center}
\renewcommand{\arraystretch}{1.2}
\begin{tabular}{lcc}
\multicolumn{3}{c}{\scriptsize TABLE 3}\\
\multicolumn{3}{c}{\scriptsize MID-INFRARED FLUX MEASUREMENTS}\\
\hline
\hline
%\vspace{0.2cm}
Name & N flux (Jy) & IHW18 flux (Jy)\\
\hline
49 Ceti & 0.25$\pm$20\% & \\
51 Oph & & 8.95$\pm$20\%\\
HD 35187\tablenotemark{a} & 3.40$\pm$10\% & 5.80$\pm$10\%\\
HD 38678 & 1.80$\pm$10\% & \\
HD 74956 & 6.46$\pm$10\% & 2.93$\pm$10\%\\
HD 102647 & 5.30$\pm$10\% & 1.50$\pm$10\%\\
HD 135344 & 1.29$\pm$10\% & 2.59$\pm$10\%\\
HD 143006 & 0.84$\pm$10\% & 1.85$\pm$10\%\\
HD 155826 & 2.78$\pm$10\% & 3.90$\pm$10\%\\ 
HD 163296 & 20.70$\pm$10\% & 16.77$\pm$20\%\\
HD 169142 & 2.37$\pm$10\% & 7.86$\pm$10\%\\

\hline
\end{tabular}
\end{center}
\tablenotetext{a}{\scriptsize The total flux for the binary.}
\end{scriptsize}
\end{table}

\clearpage

\begin{table}
\begin{scriptsize}
\begin{center}
\renewcommand{\arraystretch}{1.2}
\begin{tabular}{lccccc}
\multicolumn{6}{c}{\scriptsize TABLE 4}\\
\multicolumn{6}{c}{\scriptsize DERIVED PROPERTIES OF DUST DISKS}\\
\hline
\hline
Name & $\tau$ & $T_{dust}$ & $D (p=0)$ & $D (p=1)$ & Radius limit\\
\hline
49 Ceti	  &	$1.1\times 10^{-3}$ &  70K &92 AU& 1065 AU& $\sim$90 AU\\
51 Oph	  &	$2.8\times 10^{-2}$ & 500K &2 AU & 10 AU & $\lesssim$65 AU\\
HD 35187 &	$1.6\times 10^{-1}$ & 150K& 18 AU& 138 AU& $\lesssim$200 AU\\
HD 38678  &	$7.1\times 10^{-5}$ & 140K & 20 AU & 164 AU & $\lesssim$30 AU\\
HD 74956  &	$8.6\times 10^{-6}$ & 140K & 23 AU& 188 AU & $\lesssim$25 AU\\
HD 102647  &	$1.9\times 10^{-5}$ & 120K & 25 AU & 210 AU & $\lesssim$20 AU\\
HD 135344 & 	$2.7\times 10^{-1}$ & 110K & 9 AU& 64 AU& $\lesssim$60 AU\\
HD 143006&	$3.7\times 10^{-1}$& 120K& 5 AU& 35 AU& $\lesssim$50 AU\\
HD 155826  &	$6.5\times 10^{-4}$ & 110K &9 AU& 68 AU& $\lesssim$20 AU\\
HD 163296&	$9.8\times 10^{-2}$& 120K & 32 AU& 277 AU& $\lesssim$60 AU\\
HD 169142&	$2.5\times 10^{-1}$&  95K & 30 AU& 277 AU& $\lesssim$150 AU\\
\hline
\end{tabular}
\end{center}
\end{scriptsize}
\end{table}

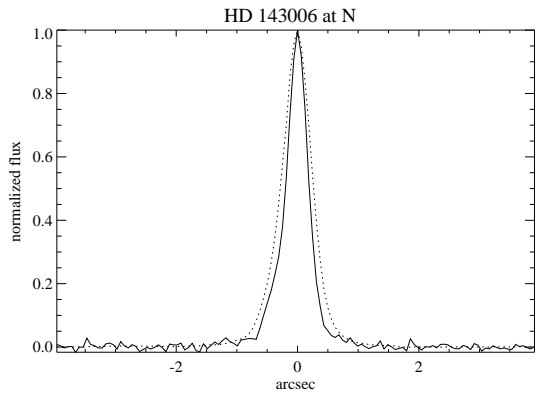
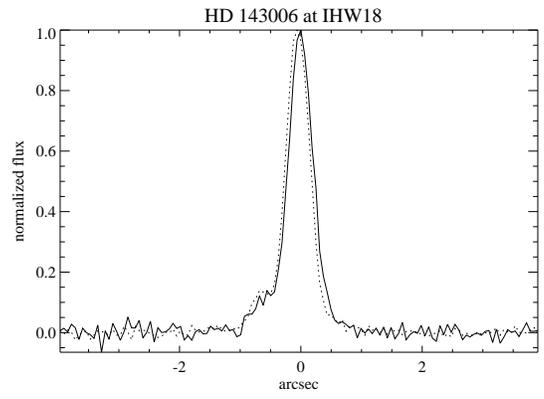
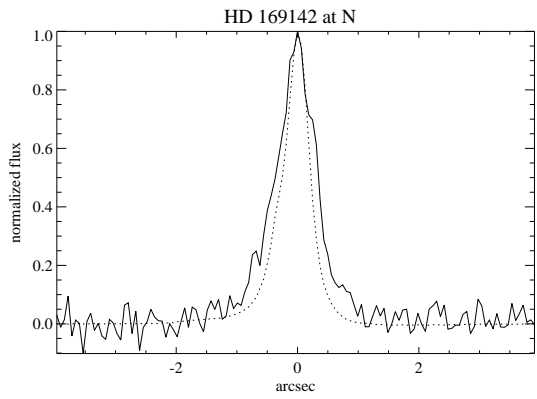
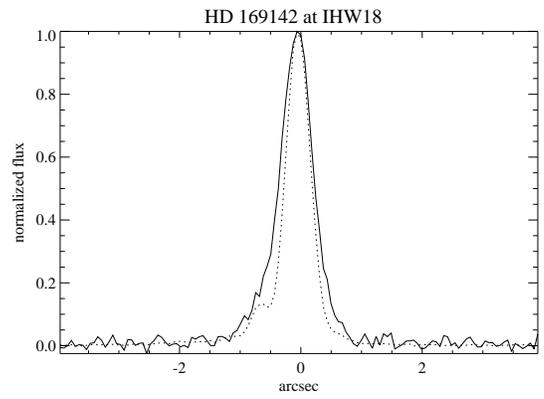
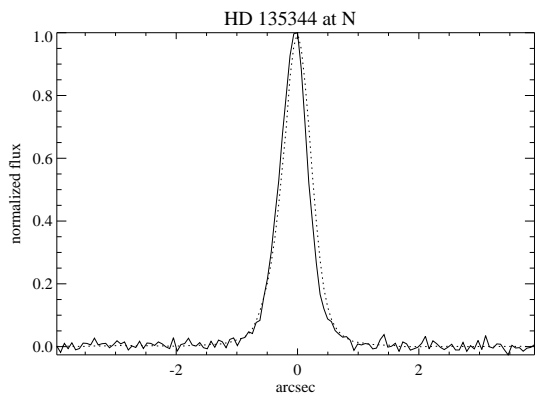
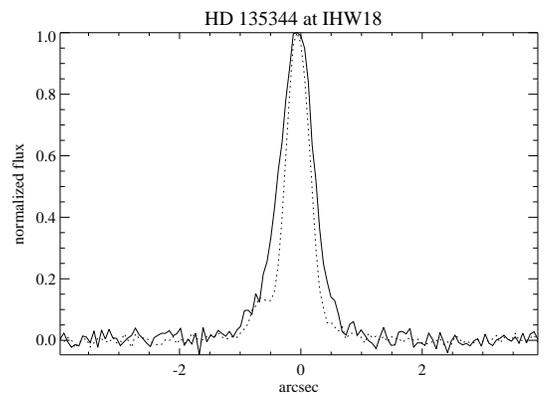
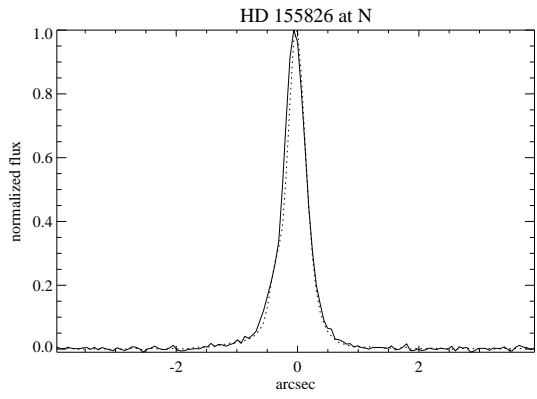
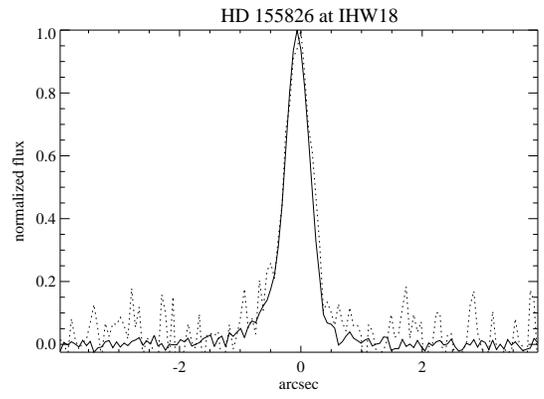
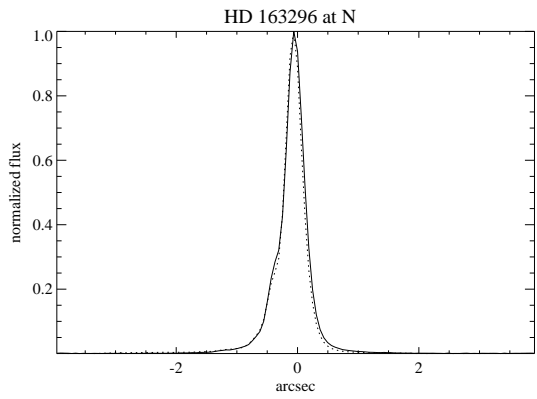
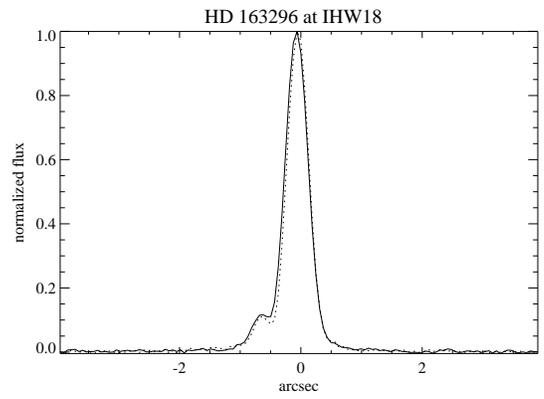
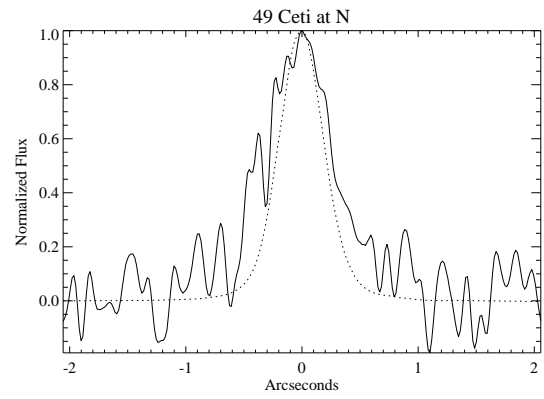
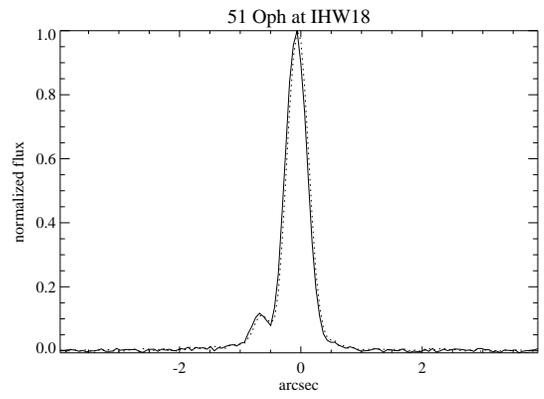
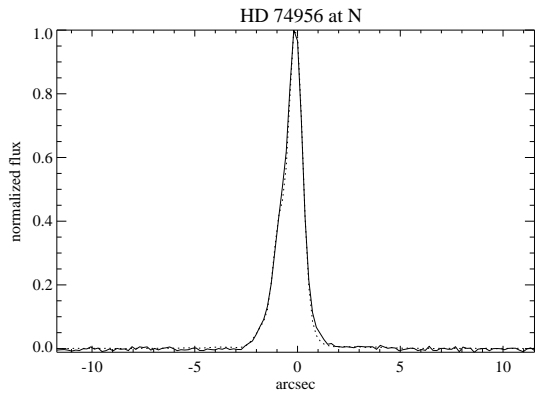
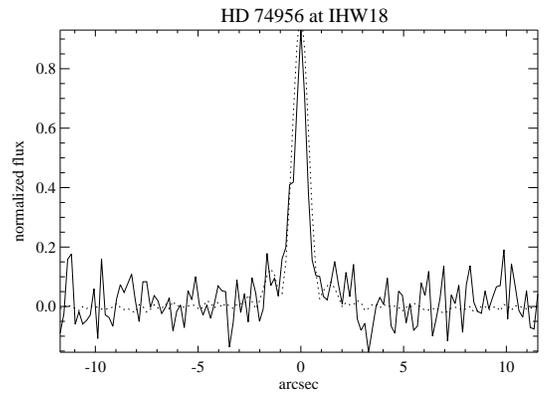
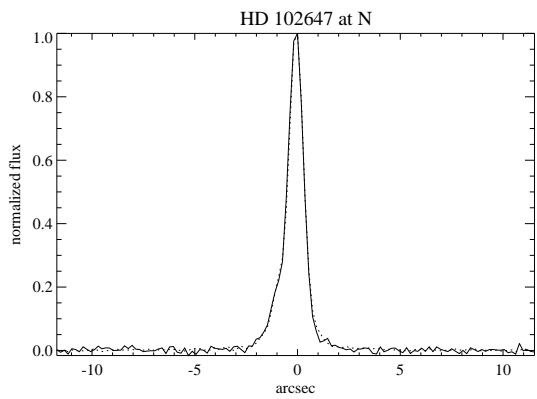
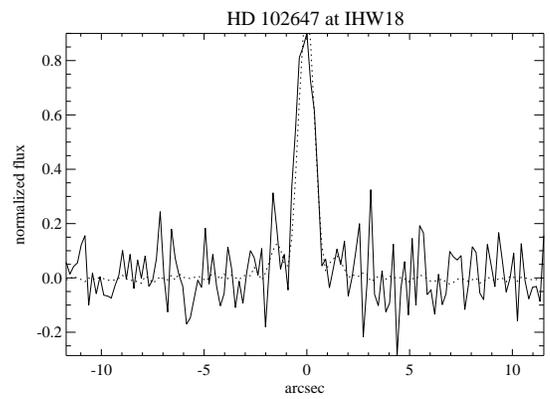
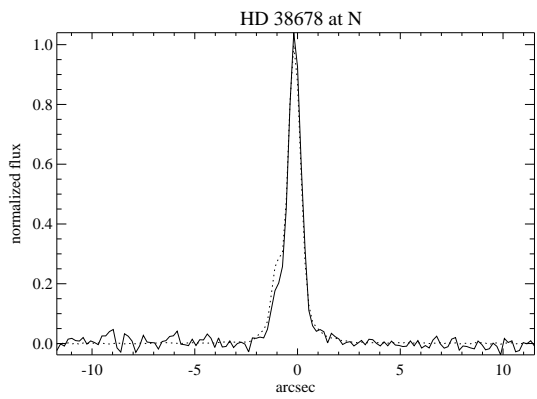
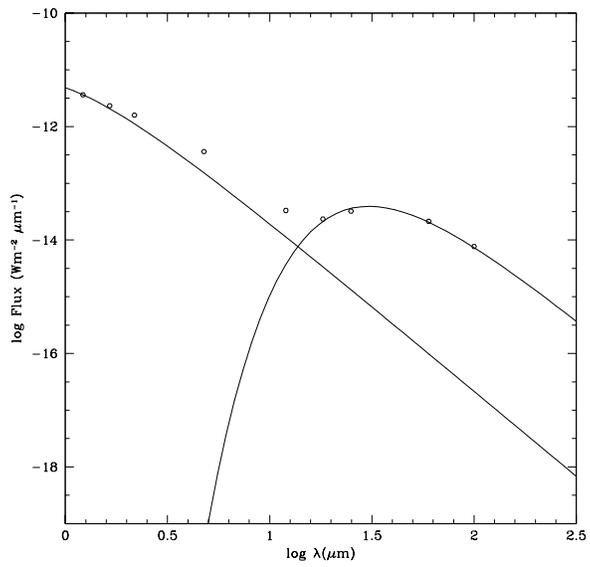
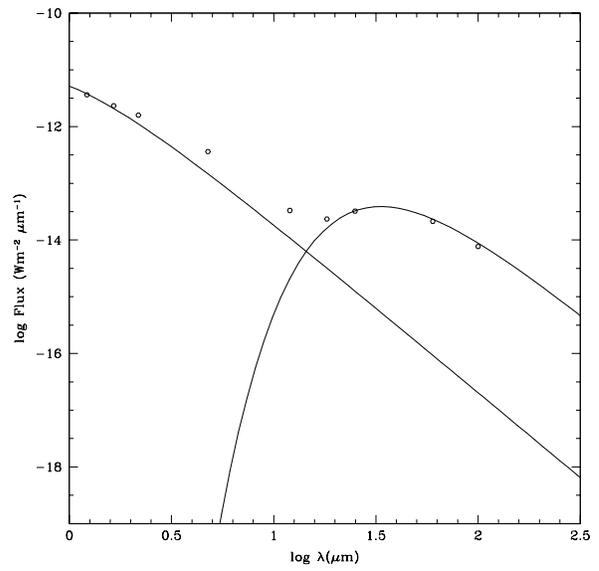
\end{document}